\documentclass[aps,prl,twocolumn,groupedaddress,showpacs]{revtex4}
\usepackage{graphics,graphicx}
\usepackage{color}
\usepackage{wrapfig}
\usepackage{enumerate}
\usepackage{amssymb,amsmath}
\usepackage{bookmath}
\usepackage{bm}

\newcommand{\req}[1]{Eq.~(\ref{#1})}

\newcommand{\rfig}[1]{Fig.~\ref{#1}}

\begin{document}
\unitlength = 1mm
\title{Nodes vs. minima in the energy gap of iron-pnictides from field-induced anisotropy}

\author{A.~B.~Vorontsov}
\affiliation{Department of Physics,
             Montana State University, Bozeman, Montana, 59717, USA}
\author{I.~Vekhter}
\affiliation{Department of Physics and Astronomy,
             Louisiana State University, Baton Rouge, Louisiana, 70803, USA}

\date{\today}

\begin{abstract}
We develop the formalism for computing the
oscillations of the specific heat and thermal transport under rotated magnetic field
in multiband superconductors with anisotropic gap and apply it to iron-based materials.
We show that these oscillations change sign at low temperatures and fields,
which strongly influences the experimental conclusions about the gap structure.
We find that recent measurements of the specific heat oscillations indicate that
the iron-based superconductors possess an anisotropic gap with deep minima or nodes
close to the line connecting electron and hole pockets.
We predict the behavior of the thermal
conductivity that will help distinguish between these cases.
\end{abstract}
\maketitle

Discovery of superconductivity in iron-pnictides
re-energized the effort to understand the
properties of the paired state in correlated electron systems.
Materials containing FeAs~\cite{Kamihara:2008sc1111,Rotter:2008scK122}
or Fe(Se,S,Te)~\cite{Hsu:2008sc11,Subedi:2008dft11} layers
are quasi-two-dimensional, and most are antiferromagnetic (AFM) at
stoichiometry, which led to early comparisons to the cuprate superconductors (SCs). 
Unlike cuprates, the pnictides
remain metallic~\cite{Kamihara:2008sc1111,Rotter:2008scK122,Hsu:2008sc11,Subedi:2008dft11,Klauss:2008commSDW,Eremin:2010metal,Ishida:2009review}
in the AFM state, suggesting that the itinerant correlated picture
is more appropriate. This view is also
supported by the agreement between  the band structure
calculations and the measured Fermi surface~\cite{MazinSchmalian,Lu:2008arpesP,Liu:2009review,Singh:2009review,Analytis:2009itinerant}.

The complication is that all five $d$-orbitals of Fe ions contribute to the
density of states (DOS) close to the Fermi level resulting in multiple Fermi
surface (FS) sheets. The complete description of these materials includes two
to three hole bands at the center ($\Gamma$-point) of the Brillouin Zone (BZ)
and two electron bands in the corner of the BZ,
hereafter referred to as $M$-point.
This multiband nature is essential for the ongoing debate about the
structure of the superconducting order parameter in
iron-based SCs.

Superconductivity in pnictides is likely 
due to the magnetically assisted electron scattering
between the nearly-nested hole (h) and electron (e) FS sheets~\cite{MazinSchmalian},
leading to a so-called $s^\pm$ state, with
both pockets fully gapped, and $\Delta_e=-\Delta_h$.
\cite{Mazin:2008splus,Chubukov:2008rg,Wang:2009frg}
Detailed description of spin fluctuations and intra-band Coulomb scattering
favors anisotropic gaps: only on the electron sheets for the $A_{1g}$
(extended $s$-wave) representation,
\cite{Kuroki:2008band5,Seo:2008splus,Maier:2009,Chubukov:2009nodes,Thomale:2009nodes},
and on all FS sheets for the gap of $B_{1g}$ ($d$-wave) character~\cite{Graser:2009degenr,Goswami:2009degenr}.
The latter gap shape is unlikely since ARPES measurements see nearly uniform gaps on hole FS.
  \cite{Liu:2009review,Ding:2008arpesK122,Liu:2008arpesK122,Nakayama:2009arpesK122,Evtushinsky:2009scgapK122,Lin:2008arpesK122,Kondo:2008arpesNd1111}

The magnitude of the anisotropic component in the $A_{1g}$ state
depends on the values of the interaction parameters and hence
is material-dependent.
Thus, possibilities range from isotropic $\Delta_e$,
to a gap with deep minima on the electronic FS along $\Gamma$-$M$ line,
to a state with a pair of {\it ``accidental''}
nodes near this line~\cite{Chubukov:2009nodes,Maier:2009,Kuroki:2009switch}.
In the unlikely case of the dominant anisotropic component, the nodes move to
positions along the sides of the crystallographic BZ.
Signatures of low-energy excitations were found in
Co-doped Ba(FeAs)$_2$~\cite{Tanatar:2010doping},
LaFePO\cite{Fletcher:2009nodesP,Hicks:2009nodesP},
BaFe$_2$(As$_{1-x}$P$_x$)$_2$~\cite{Hashimoto:2009univers,Yamashita:2009thcnP,Nakai:2010nodesP}
and Fe(Se,Te)~\cite{Bendele:2010nodes11} materials.
However, the detailed gap structure of the pnictides, including
the location of the possible nodes on the electron sheet,
still needs to be unambiguously determined.

Oscillations of the thermodynamic and transport coefficients in
SCs with anisotropic gap
as a function of the relative orientation of the magnetic field and the nodal (or
quasi-nodal) directions
~\cite{Vekhter:1999dos,Vekhter:2001dTD,Vorontsov:2006invC,Vorontsov:2007prb1,Vorontsov:2007prb2}
are extensively used to determine the position of the \emph{symmetry-enforced} gap nodes
~\cite{Matsuda:2006review,TSakakibara:2007}.
The key prediction of the inversion of the anisotropy~\cite{Vorontsov:2006invC},
the switch from the minima to maxima for the field along the direction of the smallest gap,
was recently confirmed~\cite{An:2010invers}.
A similar test was suggested for pnictides in Ref.~\cite{Graser:2008doppler},
and very recent measurements of the specific heat in the vortex state
of Fe(Se,Te)~\cite{Zeng:2010anisotr11}
were interpreted as leading to a surprising
conclusion that the nodes of the gap are along the principal
directions in the Brillouin Zone. This stimulated our study.

In this Letter we develop the formalism for computing the properties of the
vortex state of multiband two-dimensional (2D) superconductors under an
in-plane magnetic field.
We specifically address the states without symmetry-enforced nodes,
such as pnictides.~\cite{footnote}
We analyze the behavior
of the specific heat, ($C$), and the electronic thermal conductivity, ($\kappa$), focusing
on the regime where the inversion of the oscillations occurs,
and compare it with data in Ref.~\cite{Zeng:2010anisotr11}.
Accounting for the inversion we find that,
contrary to the conclusions of that
paper, the results are most consistent with either deep minima or nodes
close to the $\Gamma$-$M$ direction. We predict the evolution of the
$C$ and $\kappa$ 
as a
function of the field direction for different temperatures and fields.

\begin{figure}[t]
\centerline{\includegraphics[width=\linewidth]{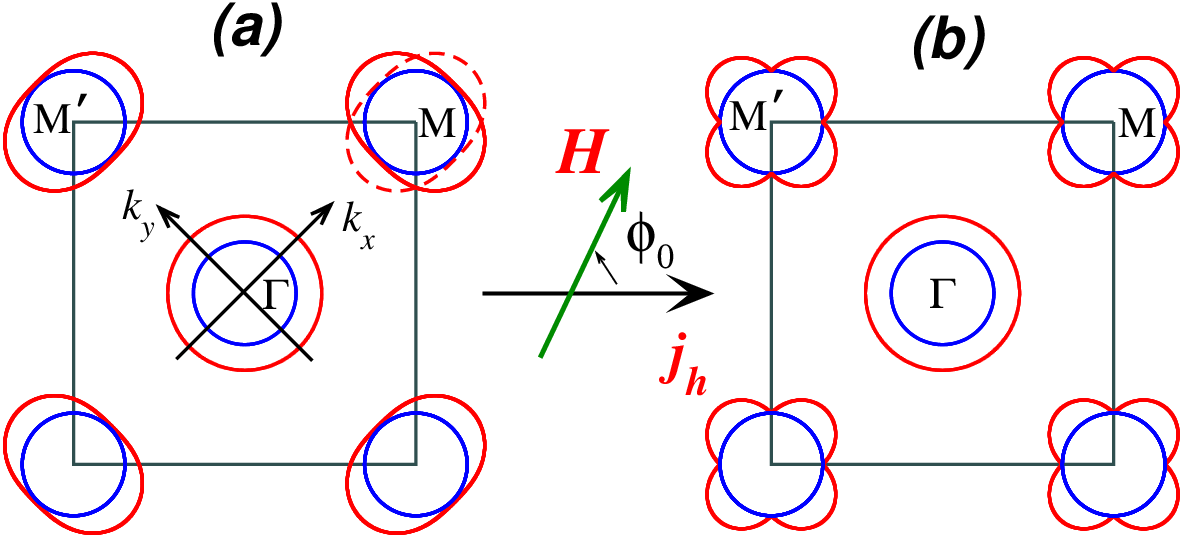}}
\caption{(Color online)
Fermi surfaces and energy gaps in the two band model of iron-pnictides.
The true structure is obtained by folding the FS sheets (blue) and SC gaps (red)
along the sides of the square (full and dashed lines at $M$). 
The hole FS around $\Gamma$ has an isotropic gap.
The $A_{1g}$ gap on the electronic FS may be isotropic,
have deep minima along $\Gamma$-$M$ line [panel (a), $r=0.45$ in \req{eq:Ye}]
or nodes along the [100] as in panel (b),  $r=1$.
We take 
the heat current along [100], as shown.
}
\label{fig:FS}
\end{figure}

We consider a superconducting gap
with the basis function in the $A_{1g}$ representation,
$\cY(k_x, k_y) = a + b (\cos k_x + \cos k_y)$,
where $k_x$ and $k_y$ belong to the unfolded Brillouin Zone, \rfig{fig:FS}(a), and
the ratio $a/b$ determines the anisotropy of the gap.
In that scheme a circular hole sheet is
centered around $\Gamma$ point, and the
electron sheets are centered around $M$ point and its equivalents.
This choice of
$\cY(k_x, k_y)$ gives nearly isotropic order parameter on the hole
FS, $\Delta_h(\phi) = \Delta_1$, 
and, generally, an anisotropic gap on the electron FS,
$\Delta_e(\phi) = \Delta_2 \cY_e(\phi)$, 
\be
\cY_e(\phi) = (1-r) \mp r \sin 2\phi \,, \quad \mbox{$\mp$ for $M$,$M'$}\,,
\label{eq:Ye}
\ee
where $\phi$ is the angle 
from the [100] direction,
see \rfig{fig:FS}. 
The gap has either minima
along $\Gamma$-$M$ line ($r<0.5$, \rfig{fig:FS}a)
or nodes close to $\Gamma$-$M$ line for $r\gtrsim 0.5$,
which approach [100] and [010] directions for $r=1$, \rfig{fig:FS}b.
We consider these cases below.

We compute the field-dependent specific heat and electronic thermal
conductivity
by solving the equation for the quasiclassical
Green's function $\whg$ in the particle-hole (Nambu) space, as in Refs.~\cite{Vorontsov:2007prb1,Vorontsov:2007prb2}.
To treat multiple bands we introduce
$\whg_{n=1,2}(\vR, k; \vare)$, with $k$-parameter running
over the hole FS ($n=1$) twice (two equivalent hole bands),
and over the two electronic FSs ($n=2$) at points $M$ and $M'$.
In each band, given the Fermi velocity $\vv_n(k)$ and the vector potential
$\vA(\vR)$ of the magnetic field, we solve the transport equation for $\whg_n$ at energy $\vare$,
\be
\left[ \left( \vare + {e\over c} \vv_n \vA(\vR) \right) \, \widehat{\tau}_3
- \whDelta_n  - \whs_{n} , \whg_n\right] + i\vv_n \cdot \gradR \; \whg_n = 0 \,,
\label{eq:eil}
\ee
subject to the normalization condition $\whg_n^2=-\pi^2 \widehat{1}$.
Here $\widehat{\tau}_3$ is the Pauli matrix, and the spatial dependence of the order parameter is
that of the Abrikosov vortex lattice,
\begin{equation}
  \Delta_n (\vR, k) = \Delta_n(k) \sum_{l_y} C_{l_y}
  \frac{{e^{il_y\widetilde y}}}{\sqrt{ \Lambda}}
  \widetilde\Phi_0\left( {\widetilde x-\Lambda^2
    l_y\over \Lambda } \right)
    \,,
    \nonumber
\end{equation}
where $\widetilde\Phi_0(z)$ is the ground state wave function of a harmonic oscillator,
$\widetilde x$ and $\widetilde y$ are in the plane normal to the field,
and the magnetic length $\Lambda^2 = \hbar c/2e H$. 
Two bands are mixed through the self-consistency on
$\Delta_n(\vR,k) = T \sum_{\vare_m, k', n'} V_{n,n'}(k,k') f_{n'}(\vR, k'; \vare_m)$,
where $f_{n}$, the Gor'kov pairing amplitude, is the off-diagonal component of $\whg_n$;
and on the impurity self-energy, $\whs_{n}$, which is determined in the
self-consistent $t$-matrix approximation for two bands \cite{Mishra:2009thcn},
so that $\whs(\vR; \vare)= n_{imp} \widehat{t}_s(\vR; \vare)$.
We take a negative inter-band pair hopping, $V_{12}(k,k') = -|V| \cY(k) \cY(k')$
which leads to the opposite signs of gaps $\Delta_1$ and $\Delta_2$.

We employ the extended Brandt-Pesch-Tewordt (BPT) approximation
where the diagonal components of $\whg_n$, the propagators $g_n$ and $\bar g_n$,
are replaced by their
spatial average, while the full dependence of $f_n$ is kept.
This approach is described
and justified in Refs.~\cite{AHoughton:1998,HKusunose:2004,Vorontsov:2007prb1},
and the results obtained from the
self-consistent solution of the quasiclassical equations in a single band
agree with those
obtained using our method nearly perfectly.~\cite{Hiragi:2010para}

Once the Green's function and the self-energies are determined,
we find the {low-temperature} specific heat from
\be
    \frac{C(T,\vH)}{T} = \int\limits_{-\infty}^\infty
    	\frac{d\vare}{T} \frac{\vare^2}{ 4 T^2}
	    \cosh^{-2}\frac{\vare}{2T}\sum_{n=1,2} N_n(\vare,T,\vH) \,,
    \ee
where
$N_n(\vare,T,\vH)= - \frac{1}{\pi} \langle\Im g_n(\vare,k)\rangle_{FS}$,
is the DOS in each band
and the angular brackets denote averaging over the corresponding FS. 
Similarly,
$\kappa_{xx} = \kappa_{1,xx} + \kappa_{2,xx}$, where
each Fermi surface contributes
\begin{eqnarray}
\frac{\kappa_{n,xx}}{T} &=&
\int\limits^{+\infty}_{-\infty} \; \frac{d\vare}{T}
\frac{\vare^2}{2 T^2} \cosh^{-2}\frac{\vare}{2T}
\\
\nonumber &\times& \left\langle v_{n,x}^2  \, N_n(T, \bm H; k, \vare) \;
\tau_{H,{n}}(T, H; k,\vare) \right\rangle_{FS} \,,
\end{eqnarray}
and the transport scattering rate in each band is
\cite{Vekhter:1999bpt,Vorontsov:2007prb2}
    \be
    \frac{1}{2\tau_{H,{n}}} =
    - \Im \Sigma_n^R + \sqrt{\pi}{2 \Lambda \over |\tilde{v}_n^\perp|}
    \frac{\Im[g_n^R \, W(2\tilde{\vare}\Lambda/|\tilde{v}_n^\perp|)]}
    {\Im \, g_n^R} |\Delta_n(k)|^2
    \,.
    \label{eq:tauH}
    \ee
Here $R$ indexes a retarded function, $\Sigma_n=(\whs_n)_{11}$, and $\tilde{v}_n^\perp$ is the
component of the Fermi velocity normal to $\bm H$.

\begin{figure}[t]
\centerline{\includegraphics[width=\linewidth]{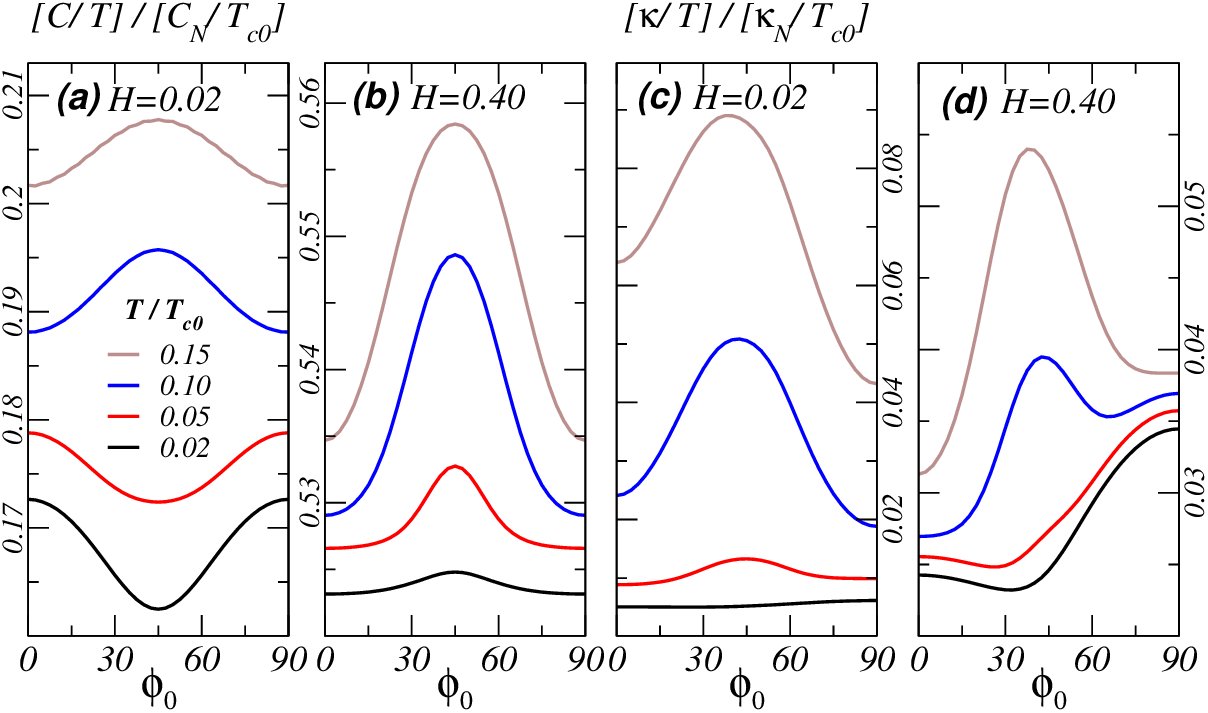}}
\caption{(Color online)
The anisotropy of the heat capacity (a,b) and thermal conductivity (c,d)
for a gap with deep minima along $\Gamma$-$M$ line ($\phi_0 = 45^\circ$),
as depicted in \rfig{fig:FS}(a).
Except for the low-$T$, low-$H$ regime, the minima
in the gap are marked by the maximum of $C$.
For heat transport, note the same trend, and almost complete
absence of the two-fold anisotropy in $\kappa(\phi_0)$
at low $H$ and $T \gtrsim 0.05 T_{c0}$.
}
\label{fig:deepmin}
\end{figure}

{\it Results.}
We take the e- and h- FSs to be cylinders of the same size, with
$|\tilde{v}_n({\bm k})|=v_f$.
In the supplementary material we show that our conclusions
are robust against modifications of the band structure,
and only for the particular case of high curvature of the electronic Fermi
surface along the $\Gamma$-$M$ line additional care is needed~\cite{Aux}.
The unit for the magnetic field is
$B_0 = \hbar c/ (2e \xi_0^2)$,
where $\xi_0 = v_f/2\pi T_{c0}$ is the in-plane coherence length,
and $T_{c0}$ is the transition temperature in a pure sample.
Highly anisotropic pairing
states are affected by the disorder in each band.
Hence below we present the results for the purely {\em intra}-band
impurity scattering limit, with the normal state scattering rate,
$\Gamma/2\pi T_{c0} = 0.005$,
which gives $\lesssim 5\%$ suppression of the transition temperature.
We consider
strong scatterers, phase shift $\delta=\pi/2$,  and checked that 
smaller $\delta$ and moderate inter-band scattering do not
perceptibly change our results. While
the BPT method works well for nodal superconductors,
its validity in systems with finite minimal gap $\Delta_{min}$, 
is restricted to the regime
where 
$v_f/(2\Lambda\Delta_{min})\geq 1$.~\cite{Vorontsov:2007prb1} 
For $r=0.45$ in Eq.(\ref{eq:Ye}), at the lowest field we consider, $H = 0.02 B_0$,
this ratio is about 2.
Taking $\xi_0 \sim 30 \AA$ gives $B_0 \simeq 35\,T$, so that the fields up to
$14\,T$ correspond to $H \lesssim 0.4\,B_0 \ll H_{c2}$.
The last inequality allows for
non-self-consistent calculation
of the order parameter suppression by the in-plane $\vH$,
which is known to be an excellent approximation.
~\cite{Vorontsov:2007prb1,Hiragi:2010para,Boyd:2009doppler}

\begin{figure}[t]
\centerline{\includegraphics[width=\linewidth]{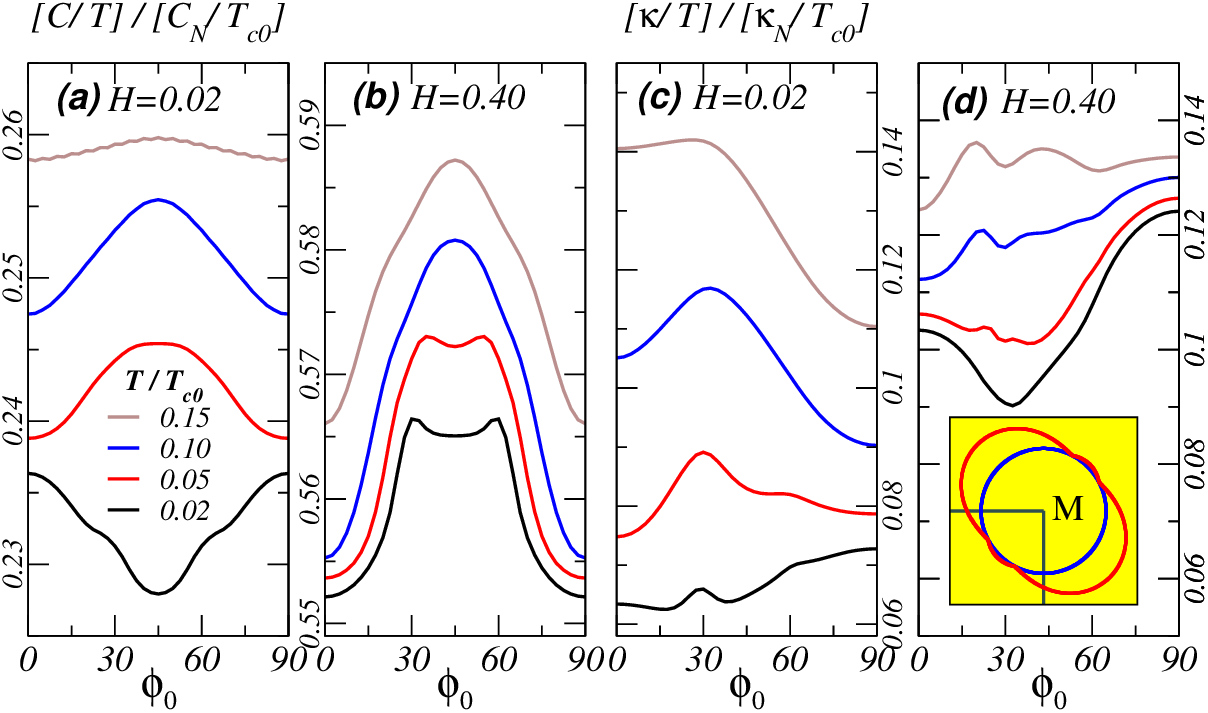}}
\caption{(Color online)
The anisotropy of $C/T$ (a,b) and $\kappa/T$ (c,d)
for $r=0.55$, gap with a pair of nodes close to the $\Gamma$-$M$ line. The overall
behavior is similar to that for
deep minima, \rfig{fig:deepmin}. The additional structure in $C/T$ for near-nodal directions
disappears already at low $T$, panels (a) and (b);
Thermal transport shows additional structure to higher $T$,
and the twofold component is dominant, panels (c) and (d).
}
\label{fig:closenodes}
\end{figure}

\begin{figure}[t]
\centerline{\includegraphics[width=\linewidth]{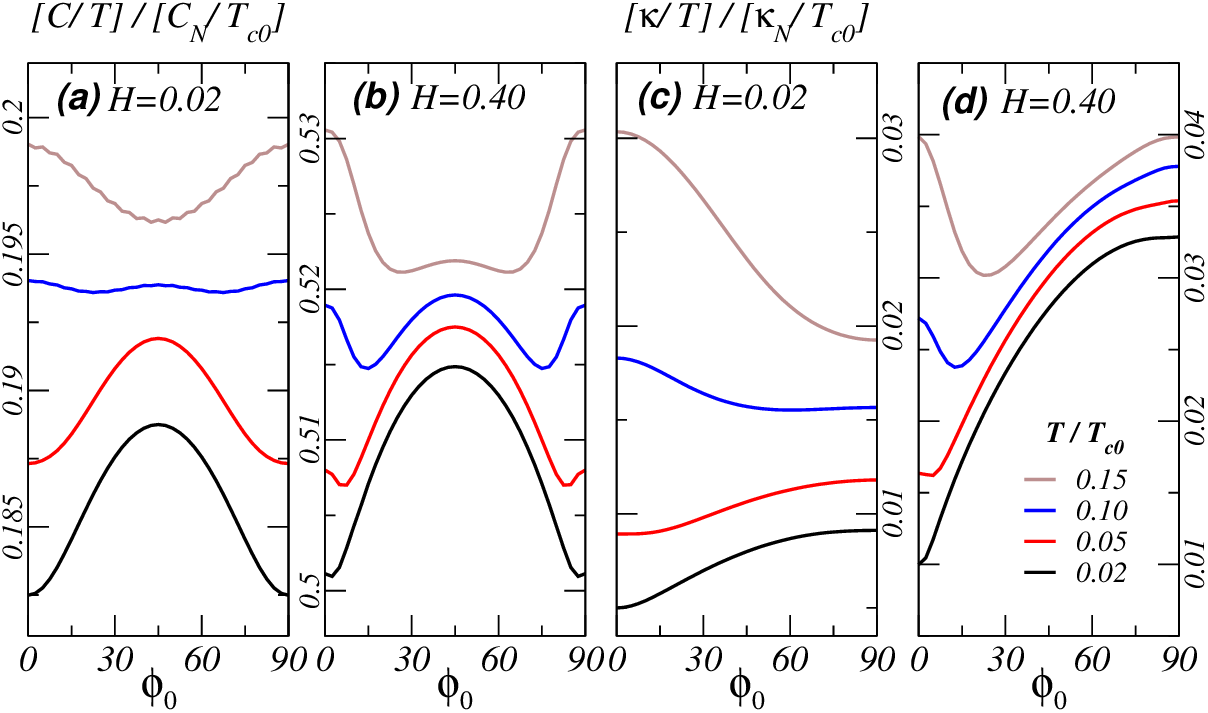}}
\caption{(Color online)
Anisotropy of $C$ and $\kappa$ for $r=1$, nodes along [100], as in
\rfig{fig:FS}(b).
The nodal directions $\phi_0=0,90^\circ$, are marked by minima of $C$
only at $T\lesssim 0.1T_c$. At higher $T$, $H$ the
angle-dependence shows a maximum of $C$ for those directions, as well as additional structure.
}
\label{fig:nodes}
\end{figure}

Figs.~(\ref{fig:deepmin})-(\ref{fig:nodes}) show representative results
for 
$C$ and $\kappa$ as a
function of the field direction at low fields for different order parameter
structures, $r=0.45,0.55,1$.
The panels capture the qualitative behavior across the $T$-$H$ phase
diagram, with only quantitative changes as we go to higher fields and
temperatures.
The $C$ and $\kappa$ profiles are slightly shifted vertically for clarity.

\rfig{fig:deepmin} and \rfig{fig:closenodes} use the gap suggested by the
majority of theoretical works, with either minima [$r=0.45$,
\rfig{fig:FS}(a)] or closely spaced nodes [$r=0.55$, inset in
\rfig{fig:closenodes}(d)] in the $\Gamma$-$M$ direction, $\phi_0=45^\circ$ in our
notation. The key feature is the inversion of the specific
heat oscillations as the temperature is raised. At the lowest $T,\,H$
the minima in $C/T$ indeed occur for $\vH$ along the $\Gamma$-$M$
line, as expected from the semiclassical theory
for zero-energy DOS,\cite{Vekhter:1999dos,Vekhter:2001dTD,Graser:2008doppler}
but this regime is narrow, and at higher $T$ and $H$ it is the maxima of
the $C$-pattern that denote the minima/nodal directions. At low fields the
inversion occurs at $0.05 < T/T_{c0} < 0.1$ for deep minima,
\rfig{fig:deepmin}(a),  and at even lower $T/T_{c0} \leq 0.05$ for
the nodal case, \rfig{fig:closenodes}(a).
The same inversion appears for {\em all} $T$ 
at fields $H\geq 0.3 B_0$,
\rfig{fig:deepmin}(b) and \rfig{fig:closenodes}(b).
In Ref.~\cite{Zeng:2010anisotr11} the $C/T$ anisotropy
was measured at $T/T_c\approx 0.2$. 
At this temperature the minima in $C/T$, observed at $\phi_0 = 0, 90$ relative
to the crystallographic axes, indicate 
deep minima at $\phi_0 = 45^\circ$ or nodes close to this direction
as evident from comparison with the upper curves in
\rfig{fig:deepmin}(a,b) or \rfig{fig:closenodes}(a,b).

In contrast, the experimentally observed pattern is {\em not consistent} with the
nodes along [100] and [010]. 
\rfig{fig:nodes}(a),(b) show minima in the $C/T$ pattern for 
the field in the nodal direction, 
$\phi_0=0$, only at $T/T_{c0}\leq 0.10$;
at higher $T$ additional structure develops, followed by
the inversion  and the shift of the minima of $C(\phi_0)$ to $\phi_0=45^\circ$.
This is not what was found in Ref.~\cite{Zeng:2010anisotr11}. 

Hence possible gap structures are (a) minima along the $\Gamma$-$M$ line,
or (b) nodes close to this direction.
Distinguishing between the two
by methods sensitive to the amplitude but not the
phase of the gap is not straightforward.
While
\rfig{fig:closenodes} clearly shows additional features (absent in \rfig{fig:deepmin})
at the angles where $|\Delta_e|$ has nodes, these
features are washed out with increased temperature.
For comparison, $\kappa/T$ shows nodal features at higher $T$
(panels c,d) than $C/T$ (panels a,b).
With increased scattering this structure  smears out
and largely vanishes when the nodes are lifted by disorder~\cite{Mishra:2009lift}.

Commonly, $\kappa(\phi_0)$ is decomposed into a twofold term due to difference
in transport normal and parallel to the vortices~\cite{KMaki:1967}, and fourfold component due to
gap structure~\cite{Matsuda:2006review}, $\kappa=\kappa_0+\kappa_2\cos 2\phi+\kappa_4\cos 4\phi$.
Both $\kappa_2$ and $\kappa_4$ change sign in the $T$-$H$
plane~\cite{KMaki:1967,Vorontsov:2007prb2}.  The magnitude of 
$\kappa_2$ is similar in both \rfig{fig:deepmin} and \rfig{fig:closenodes},
however, the fourfold $\kappa_4$ is much greater for
the gap minima
than for either nodal scenario.
Observation of a large ratio $\kappa_4/\kappa_2 \gtrsim 1$
would indicate minima and not nodes in the gap.

Additional information is needed to distinguish between shallow and deep minima.
In \rfig{fig:deepmin} for $r=0.45$, ($\Delta_{min}/\Delta_{max} = 0.1$)
the  anisotropy in $C$ is $\sim 5\%$ for our choice of the FSs.
Setting  
$r=0.3$ ($\Delta_{min}/\Delta_{max} = 0.4$)
we found that this anisotropy drops below
1\%, 
and its inversion occurs at 
$0.15 \lesssim T/T_{c0} < 0.2$
for similar fields.
Also, above the inversion, the ratio $\kappa_4/\kappa_2$ is much smaller than that for $r=0.45$.
Of course, larger minimal gap should be evident from
measurements even at zero field.

{\it Conclusions.}
We developed a framework for the calculation of the anisotropy in the heat
capacity and thermal conductivity of two-band superconductors under rotated
magnetic field, presented the results for several models of pnictides with
anisotropic $A_{1g}$ (`extended $s$') gap symmetry, and compared them with
an experiment on the Fe(Se,Te) system.
We identify either minima ($\Delta_{min}/\Delta_{max} < 0.4$) or nodes in
the gap on the electronic FS along the $\Gamma$-$M$ line, 
contrary to Ref.~\cite{Zeng:2010anisotr11}.  We predict that
comparison of the fourfold and twofold term in the anisotropy of thermal
transport  will help distinguish between the two scenarios.
Experiments in a wider $T$-$H$ range 
along with the calculations based on 
realistic FS
are clearly forthcoming. Our work lays the foundation for determining
the gap structure from these measurements.

We thank A. V. Chubukov, P.~J.~Hirschfeld, Y.~Matsuda and T.~Shibauchi for
valuable discussions, and acknowledge partial support from DOE Grant DE-FG02-08ER46492
(I.~V.) and Aspen Center for Physics.

{\it Note added:} 
After our manuscript was submitted 
similar conclusions were reached independently 
in Ref.~\cite{ChubukovEremin}



\newpage
{\bf Supplement to
 ``Nodes vs. minima in the energy gap of iron-pnictides from field-induced anisotropy''}



\vspace*{5mm}
{\bf\small
We show that our main results presented in the Letter are robust against the
changes in the Fermi surface shape and the changes in relative magnitude of the
gaps on different Fermi surface sheets.
We present the anisotropy of the heat capacity, $C(\phi_0)$, for the
elliptical Fermi surfaces and for pairing interactions resulting in a
significant difference in the gap magnitudes on the hole and
the electron Fermi surface sheets.
}

In our Letter we considered a simple model for the Fermi surface (FS) of the
iron-based superconductors that consists of cylindrical Fermi surfaces both at
the center of the Brilloin Zone (hole-like) and at the $M$ and $M^\prime$
points (electron-like). While this is likely an adequate approximation to
the electronic structure of the Fe(SeTe) compound where the only experimental data
for the specific heat anisotropy are available~\cite{Zeng:2010anisotr11},
broader theoretical questions include understanding for what classes of
non-circular Fermi surfaces observed in the members of the iron superconductor
family~\cite{SBorisenko:2010} the results remain fundamentally the same, and
under what circumstances we can expect significant deviations from the behavior
outlined in our Letter.
The main point of interest is the sign of the field-induced anisotropic
component of the specific heat in the $T$-$H$ plane. An additional focus is on
the detailed angle profile of the specific heat for different Fermi surface
shapes. 
The results presented below strongly suggest that, in most situations, our main
conclusions regarding the determination of the gap structure remain valid,
although the magnitude of the oscillations and detailed angle-dependence are
sensitive to the variations of the Fermi surface. We explicitly identify the
cases where we can expect the results to be substantially modified compared to
our model predictions and be strongly dependent on the band structure for a
specific material.

There are three effects that we need to address. The first is the weak
isotropic corrugation of the Fermi surface along the $c$-axis. In our previous work for a
single band system with vertical line nodes we showed that such corrugation has
a minimal effect on the location of the inversion line
~\cite{Vorontsov:2007prb1add}. The same holds true for the current model with
either nodes or minima. More complex features may be expected if there is
substantial modulation of the gap along the $c$ direction: This is conjectured
to happen in some of the compounds~\cite{SGraser:2010,JReid:2010}, but at
present there is no universal agreement on what form the gap may take, and we
postpone the analysis of this issue until a more detailed experimental and
theoretical analysis narrows the possible gap structures. In the absence of a
strong gap variation along the $c$ direction our results hold, and we can,
without loss of generality, consider a two-dimensional model.

\begin{figure}[b]
\centerline{\includegraphics[width=\linewidth]{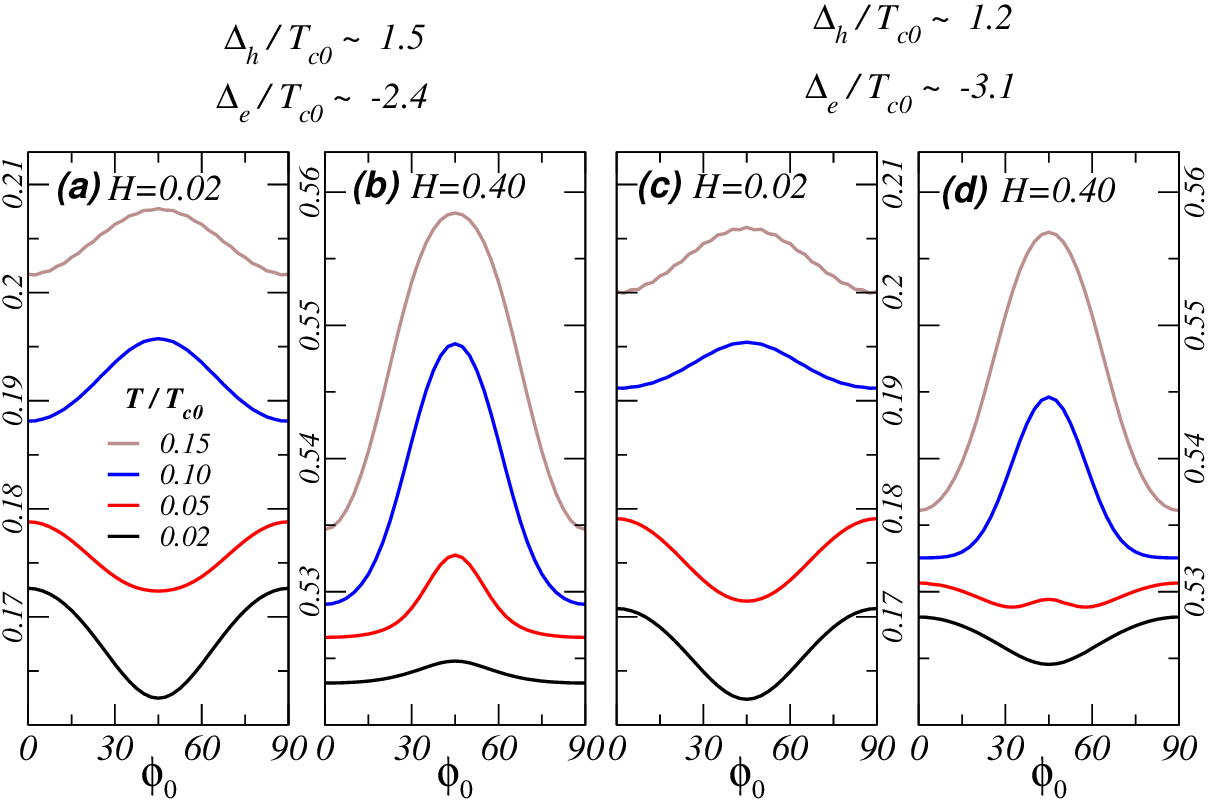}}
\caption{(Color online)
Heat capacity anisotropy for different amplitudes 
of the gaps on hole and electron FSs.
Panels (a,b):  low-$T$, zero-$H$ gap amplitudes
$\Delta_h = 1.57 \, T_{c0}$, $\Delta_e = -2.48 \, T_{c0}$.
Panel (c,d): much greater gap amplitude on the electron FS:
$\Delta_h = 1.25 \, T_{c0}$, $\Delta_e = -3.12 \, T_{c0}$.
}
\label{fig:gap}
\end{figure}

The second aspect that can potentially affect the variation of the specific heat
with angle is the difference in the gap magnitude on the electron and the hole
sheets of the Fermi surface. Experiments indicate that one hole sheet may
have a gap with the magnitude only half that on the other hole and electron Fermi surface
sheets~\cite{Nakayama:2009arpesK122add,Ding:2008arpesK122add}. 
We model an extreme scenario when both hole pockets have a reduced gap. 
Fig.~\ref{fig:gap}
shows that our results are very weakly dependent on the ratio of the maximal
gaps: only around $T/T_c\lesssim 0.05$ and at moderately high fields there is a
difference in the profiles. Over most of the parameter range it is only the
anisotropy of the gap that matters, and therefore changing the overall scale
factor leads to a slight change in the number of available quasiparticle
states, but not to a change in the pattern of the dependence on the direction
of the field.

\begin{figure}[t]
\centerline{\includegraphics[width=\linewidth]{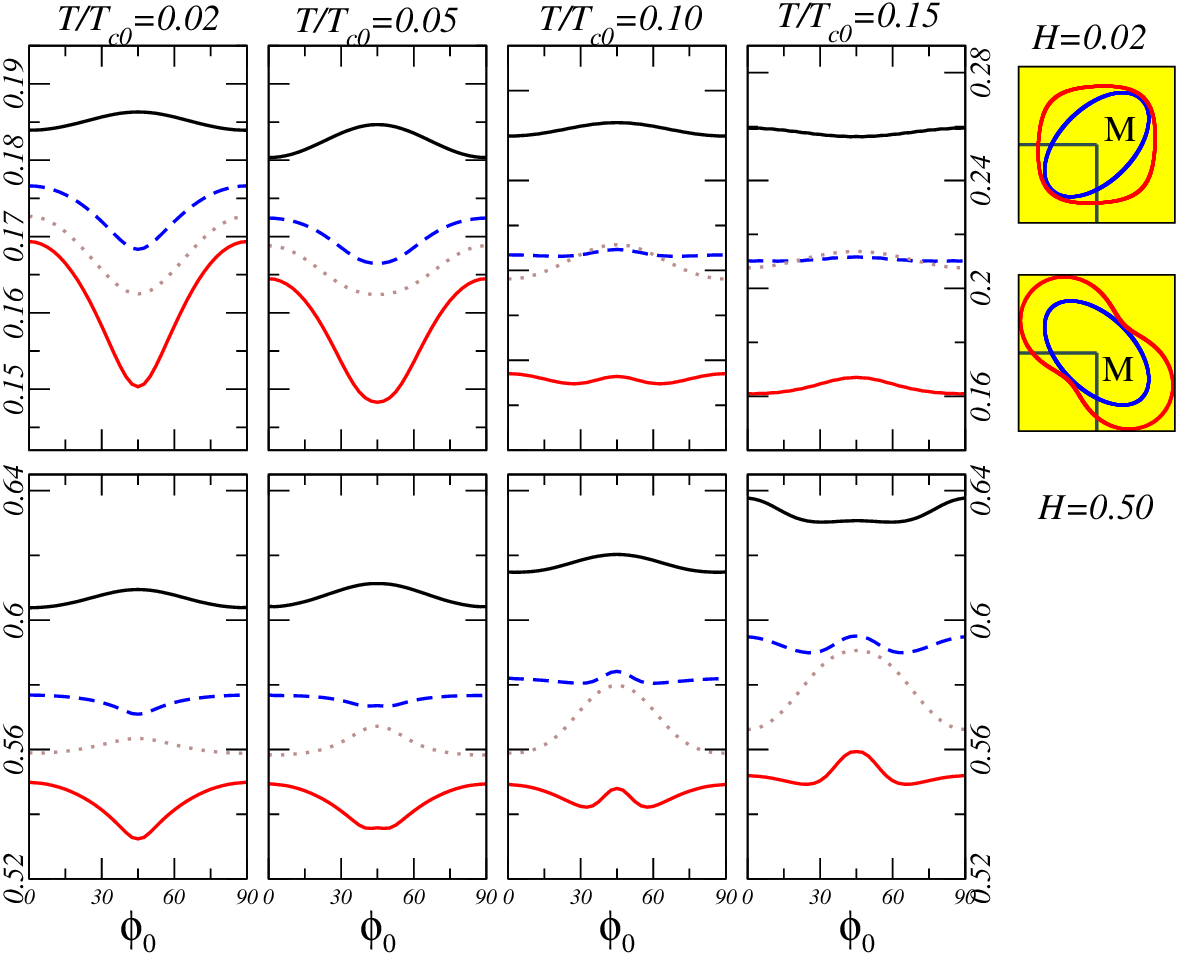}}
\caption{(Color online)
The anisotropy of the heat capacity for an elliptical electronic FS with the
aspect ratio of 2 and the gap with deep minima for different fields (upper
panels $H=0.02B_0$, lower panels $H=0.5B_0$) and temperatures from
$T/T_{c0}=0.02$ to $T/T_{c0}=0.15$. The contribution from each Fermi surface
(upper and lower solid lines in each panel) depends on whether the minima in the gap
are on the flat or strongly curved part of the ellipse.  Dashed line is the
average of the two Fermi surfaces.
Dotted lines are the results of calculations for a circular FS. }
\label{fig:ellipse}
\end{figure}


The third, and  potentially the most important effect is that of the electron Fermi
surface shape. When the Fermi surface is not circular, it is the combined anisotropy of
the Fermi surface and the gap that determines the angle-dependence of the
specific heat and the thermal conductivity. For the density of states at zero
energy the Fermi surface effects have been investigated for borocarbide
superconductors~\cite{MUdagawa:2005,YNagai:2007,YNagai:2009}, and for a model
single Fermi surface at finite temperatures~\cite{IVekhter:2008}.
Fig.~\ref{fig:ellipse} shows the results for the case of pnictide-like two-band systems
with the electron Fermi surfaces in the shape of an ellipse with the aspect
ratio of 2,
and the gap with deep minima, $r=0.45$ in
$\Delta_e(\phi) = \Delta_2  [(1-r) - r \sin 2\phi]$.
This is the simplest and most generic implementation of the Fermi
surface anisotropy that does not rely on the material-specific calculations.
Since the gap on the hole sheets is nearly isotropic the detailed shape of that
Fermi surface does not affect our results.

With finite ellipticity the two
previously degenerate Fermi surfaces at point $M$ become distinct as shown in
the rightmost column, so that the minima of the gap located along the
$\Gamma$-$M$ line now occur along one of the two principal axes of the ellipse,
see Fig.~\ref{fig:ellipse}.
The two scenarios correspond to the
gap minima at the sharp end of the ellipse
(the outer, ``flower''-like, FS in the folded zone,
shown by upper curves in Fig.~\ref{fig:ellipse}),
and the gap minima along the flatter part of the Fermi surface
(inner electron FS sheet, lower curves in Fig.~\ref{fig:ellipse}).
In the latter case,
the resulting profile of the variation of the heat capacity with the angle
strongly resembles that for the cylindrical FS.  Some additional
structure is seen at temperatures close to the inversion line, and the
inversion itself at low $T$ is pushed to slightly higher fields.

In contrast, when the minima/nodes are in the regions of the Fermi surface with
high curvature, the inversion of the angle-dependent oscillation pattern occurs
at very low values of $T$ and $H$, in agreement with Ref.~\cite{IVekhter:2008}.
Left panel of Fig.~\ref{fig:ellipse} demonstrates that this inversion already
occurred even at low field for $T=0.02T_c$. In this case small changes of Fermi
surface curvature around the loci of the gap minima do affect the
angle-dependence, and therefore for systems where this curvature is large at
the intersection with $\Gamma$-$M$ line, a detailed material-specific
calculation may be needed.

Thermal conductivity anisotropy exhibits similar changes. 
For the gap with deep minima at the flat parts of the ellipse, 
the amplitude of the fourfold term relative to the twofold component is large;  
while for the gap minima at FS regions with high curvature the twofold 
component dominates. 

In some of the pnictide materials, such as 122 As-P series, the
electronic Fermi surface is closely approximated by the ellipse in the basal
plane that undergoes 90$^\circ$ rotation as a function of the $c$-axis wave
vector~\cite{HShishido:2010}. In this situation the measured anisotropy is
essentially the average of the contributions from the two possibilities
discussed above.  This is also plotted in Fig.~\ref{fig:ellipse}, and direct
comparison with the cylindrical case shows overall similarity.  There is a more
complex angle-dependent structure at some temperatures and fields, and at very
low temperatures ($T/T_c\leq 0.05$) the higher inversion field gives a narrow
range where the oscillatory contributions have opposite sign for the circular
and averaged elliptical cross-section, see lower left panel. However, by the
experimentally relevant temperature of $T\approx 0.2T_c$ the inversion already
occurred for both cases. Therefore our main finding that the measurements at
and around this temperature indicate the minima or the nodes along the
$\Gamma$-$M$ direction remains robust in this case.

Importantly, we checked that for the elliptical electron Fermi
surfaces and nodal gap the profiles of the thermal conductivity and the
specific heat as a function of the field angle exhibit complex features
qualitatively similar to those discussed in our Letter and not seen in
experiment. 

Our conclusion therefore is that the main features of our results presented in
our Letter remain valid in most situations when changes in the Fermi surface
shape, but not its topology, are taken into account. Weak three-dimensionality
or the Fermi surface and the variation in the relative gap amplitude between
the electron and the hole sheets essentially do not affect the angle-dependence
of the specific heat. The change in the Fermi surface shape has a greater
effect, and detailed analysis is needed for the materials where the electron
Fermi surface has high curvature at the intersection with $\Gamma$-$M$ line for
most values of the $c$-axis wave vector. Without this caveat we find, however,
that within a wide range of parameters the inversion of the net measured
anisotropy of the specific heat still occurs below $T=0.2T_c$, and therefore
the experimental results of Ref.~\cite{Zeng:2010anisotr11add} are indicative of
the minima or the nodes of the gap along the $\Gamma$-$M$ line.

Of course, in some members of the large family of iron-based superconductors
the Fermi surface may have not just a different curvature but an
altogether different topology: this may happen with severe overdoping when one
of the sheets disappears altogether ~\cite{Sato:2009p589} or when additional
order leads to a reconstruction of the Fermi surface as suggested, for example,
in Ref.~\cite{Zabolotnyy:2009eorder}. In these specific cases a separate
investigation of the resulting pattern is probably warranted, but our results
are generally applicable to many compounds where the Fermi surface consists of
the hole sheets near the $\Gamma$ and (perhaps anisotropic) electron sheets
near the $M$ (or $X$) points in the Brillouin Zone.


\begin{thebibliography}{55}
\expandafter\ifx\csname natexlab\endcsname\relax\def\natexlab#1{#1}\fi
\expandafter\ifx\csname bibnamefont\endcsname\relax
  \def\bibnamefont#1{#1}\fi
\expandafter\ifx\csname bibfnamefont\endcsname\relax
  \def\bibfnamefont#1{#1}\fi
\expandafter\ifx\csname citenamefont\endcsname\relax
  \def\citenamefont#1{#1}\fi
\expandafter\ifx\csname url\endcsname\relax
  \def\url#1{\texttt{#1}}\fi
\expandafter\ifx\csname urlprefix\endcsname\relax\def\urlprefix{URL }\fi
\providecommand{\bibinfo}[2]{#2}
\providecommand{\eprint}[2][]{\url{#2}}

\bibitem{Kamihara:2008sc1111}
\bibinfo{author}{\bibfnamefont{Y.}~\bibnamefont{Kamihara}} \et,
\bibinfo{journal}{J. Am. Chem. Soc.} \textbf{\bibinfo{volume}{130}}, \bibinfo{pages}{3296}
  (\bibinfo{year}{2008}).

\bibitem{Rotter:2008scK122}
\bibinfo{author}{\bibfnamefont{M.}~\bibnamefont{Rotter}} \et,
\bibinfo{journal}{Phys. Rev. Lett.} \textbf{\bibinfo{volume}{101}},
  \bibinfo{pages}{107006} (\bibinfo{year}{2008}).

\bibitem{Hsu:2008sc11}
\bibinfo{author}{\bibfnamefont{F.-C.} \bibnamefont{Hsu}} \et,
\bibinfo{journal}{Proc. Nat. Acad. of Sci.} \textbf{\bibinfo{volume}{105}}, \bibinfo{pages}{14262}
  (\bibinfo{year}{2008}).

\bibitem{Subedi:2008dft11}
\bibinfo{author}{\bibfnamefont{A.}~\bibnamefont{Subedi}} \et,
\bibinfo{journal}{Phys.
  Rev. B} \textbf{\bibinfo{volume}{78}}, \bibinfo{pages}{134514}
  (\bibinfo{year}{2008}).

\bibitem{Klauss:2008commSDW}
\bibinfo{author}{\bibfnamefont{H.~H.} \bibnamefont{Klauss}} \et,
\bibinfo{journal}{Phys. Rev. Lett.}
  \textbf{\bibinfo{volume}{101}}, \bibinfo{pages}{077005}
  (\bibinfo{year}{2008}).

\bibitem{Eremin:2010metal}
\bibinfo{author}{\bibfnamefont{I.}~\bibnamefont{Eremin}}
  \bibnamefont{and}
  \bibinfo{author}{\bibfnamefont{A.~V.} \bibnamefont{Chubukov}},
\bibinfo{journal}{Phys. Rev. B} \textbf{\bibinfo{volume}{81}},
  \bibinfo{pages}{024511} (\bibinfo{year}{2010}).

\bibitem{Ishida:2009review}
\bibinfo{author}{\bibfnamefont{K.}~\bibnamefont{Ishida}} \et,
\bibinfo{journal}{J. Phys. Soc. Jpn.} \textbf{\bibinfo{volume}{78}},
  \bibinfo{pages}{062001} (\bibinfo{year}{2009}).

\bibitem{MazinSchmalian}
\bibinfo{author}{\bibfnamefont{I.~I.} \bibnamefont{Mazin}}
  \bibnamefont{and}
  \bibinfo{author}{\bibfnamefont{J.}~\bibnamefont{Schmalian}},
\bibinfo{journal}{Physica C} \textbf{\bibinfo{volume}{469}},
  \bibinfo{pages}{614} (\bibinfo{year}{2009}).

\bibitem{Lu:2008arpesP}
\bibinfo{author}{\bibfnamefont{D.~H.} \bibnamefont{Lu}} \et,
\bibinfo{journal}{Nature}
  \textbf{\bibinfo{volume}{455}}, \bibinfo{pages}{81} (\bibinfo{year}{2008}).

\bibitem{Liu:2009review}
\bibinfo{author}{\bibfnamefont{C.}~\bibnamefont{Liu}} \et,
\bibinfo{journal}{Physica C}
  \textbf{\bibinfo{volume}{469}}, \bibinfo{pages}{491} (\bibinfo{year}{2009}).

\bibitem{Singh:2009review}
\bibinfo{author}{\bibfnamefont{D.~J.} \bibnamefont{Singh}},
\bibinfo{journal}{Physica C} \textbf{\bibinfo{volume}{469}},
  \bibinfo{pages}{418} (\bibinfo{year}{2009}).

\bibitem{Analytis:2009itinerant}
\bibinfo{author}{\bibfnamefont{J.~G.} \bibnamefont{Analytis}} \et,
\bibinfo{journal}{Phys. Rev. B} \textbf{\bibinfo{volume}{80}},
  \bibinfo{pages}{064507} (\bibinfo{year}{2009}).

\bibitem{Mazin:2008splus}
\bibinfo{author}{\bibfnamefont{I.~I.} \bibnamefont{Mazin}} \et,
\bibinfo{journal}{Phys.
  Rev. Lett.} \textbf{\bibinfo{volume}{101}}, \bibinfo{pages}{057003}
  (\bibinfo{year}{2008}).

\bibitem{Chubukov:2008rg}
\bibinfo{author}{\bibfnamefont{A.}~\bibnamefont{Chubukov}} \et,
\bibinfo{journal}{Phys. Rev. B} \textbf{\bibinfo{volume}{78}},
  \bibinfo{pages}{134512} (\bibinfo{year}{2008}).

\bibitem{Wang:2009frg}
\bibinfo{author}{\bibfnamefont{F.}~\bibnamefont{Wang}} \et,
\bibinfo{journal}{Phys. Rev. Lett.} \textbf{\bibinfo{volume}{102}},
  \bibinfo{pages}{047005} (\bibinfo{year}{2009}).

\bibitem{Kuroki:2008band5}
\bibinfo{author}{\bibfnamefont{K.}~\bibnamefont{Kuroki}} \et,
\bibinfo{journal}{Phys. Rev. Lett.} \textbf{\bibinfo{volume}{101}},
  \bibinfo{pages}{087004} (\bibinfo{year}{2008}).

\bibitem{Seo:2008splus}
\bibinfo{author}{\bibfnamefont{K.}~\bibnamefont{Seo}} \et,
\bibinfo{journal}{Phys.
  Rev. Lett.} \textbf{\bibinfo{volume}{101}}, \bibinfo{pages}{206404}
  (\bibinfo{year}{2008}).

\bibitem{Maier:2009}
\bibinfo{author}{\bibfnamefont{T.}~\bibnamefont{Maier}} \et,
\bibinfo{journal}{Phys. Rev. B} \textbf{\bibinfo{volume}{79}},
  \bibinfo{pages}{224510} (\bibinfo{year}{2009}).

\bibitem{Chubukov:2009nodes}
\bibinfo{author}{\bibfnamefont{A.}~\bibnamefont{Chubukov}} \et,
\bibinfo{journal}{Phys. Rev. B}
  \textbf{\bibinfo{volume}{80}}, \bibinfo{pages}{140515}
  (\bibinfo{year}{2009}).

\bibitem{Thomale:2009nodes}
\bibinfo{author}{\bibfnamefont{R.}~\bibnamefont{Thomale}} \et,
\bibinfo{journal}{Phys. Rev. B}
  \textbf{\bibinfo{volume}{80}}, \bibinfo{pages}{180505}
  (\bibinfo{year}{2009}).

\bibitem{Graser:2009degenr}
\bibinfo{author}{\bibfnamefont{S.}~\bibnamefont{Graser}} \et,
\bibinfo{journal}{New J Phys}
  \textbf{\bibinfo{volume}{11}}, \bibinfo{pages}{025016}
  (\bibinfo{year}{2009}{\natexlab{a}}).

\bibitem{Goswami:2009degenr}
\bibinfo{author}{\bibfnamefont{P.}~\bibnamefont{Goswami}} \et,
\bibinfo{journal}{Europhys. Lett.}
  \textbf{\bibinfo{volume}{91}}, \bibinfo{pages}{37006} (\bibinfo{year}{2010}).

\bibitem{Ding:2008arpesK122}
\bibinfo{author}{\bibfnamefont{H.}~\bibnamefont{Ding}} \et,
\bibinfo{journal}{Europhys. Lett.}
  \textbf{\bibinfo{volume}{83}}, \bibinfo{pages}{47001} (\bibinfo{year}{2008}).

\bibitem{Liu:2008arpesK122}
\bibinfo{author}{\bibfnamefont{C.}~\bibnamefont{Liu}} \et,
\bibinfo{journal}{Phys. Rev. Lett.} \textbf{\bibinfo{volume}{101}},
  \bibinfo{pages}{177005} (\bibinfo{year}{2008}).

\bibitem{Nakayama:2009arpesK122}
\bibinfo{author}{\bibfnamefont{K.}~\bibnamefont{Nakayama}} \et,
\bibinfo{journal}{Europhys. Lett.} \textbf{\bibinfo{volume}{85}},
  \bibinfo{pages}{67002} (\bibinfo{year}{2009}).

\bibitem{Evtushinsky:2009scgapK122}
\bibinfo{author}{\bibfnamefont{D.~V.} \bibnamefont{Evtushinsky}} \et,
\bibinfo{journal}{Phys. Rev. B}
  \textbf{\bibinfo{volume}{79}}, \bibinfo{pages}{054517}
  (\bibinfo{year}{2009}).

\bibitem{Lin:2008arpesK122}
\bibinfo{author}{\bibfnamefont{Z.}~\bibnamefont{Lin}} \et,
\bibinfo{journal}{Chinese Physics Letters}
  \textbf{\bibinfo{volume}{25}}, \bibinfo{pages}{4402} (\bibinfo{year}{2008}).

\bibitem{Kondo:2008arpesNd1111}
\bibinfo{author}{\bibfnamefont{T.}~\bibnamefont{Kondo}} \et,
\bibinfo{journal}{Phys. Rev. Lett.}
  \textbf{\bibinfo{volume}{101}}, \bibinfo{pages}{147003}
  (\bibinfo{year}{2008}).

\bibitem{Kuroki:2009switch}
\bibinfo{author}{\bibfnamefont{K.}~\bibnamefont{Kuroki}} \et,
\bibinfo{journal}{Phys. Rev. B} \textbf{\bibinfo{volume}{79}},
  \bibinfo{pages}{224511} (\bibinfo{year}{2009}).

\bibitem{Tanatar:2010doping}
\bibinfo{author}{\bibfnamefont{M.~A.} \bibnamefont{Tanatar}} \et,
\bibinfo{journal}{Phys. Rev. Lett.} \textbf{\bibinfo{volume}{104}},
  \bibinfo{pages}{067002} (\bibinfo{year}{2010}).

\bibitem{Fletcher:2009nodesP}
\bibinfo{author}{\bibfnamefont{J.}~\bibnamefont{Fletcher}} \et,
\bibinfo{journal}{Phys. Rev. Lett.} \textbf{\bibinfo{volume}{102}},
  \bibinfo{pages}{147001} (\bibinfo{year}{2009}).

\bibitem{Hicks:2009nodesP}
\bibinfo{author}{\bibfnamefont{C.}~\bibnamefont{Hicks}} \et,
\bibinfo{journal}{Phys. Rev. Lett.} \textbf{\bibinfo{volume}{103}},
  \bibinfo{pages}{127003} (\bibinfo{year}{2009}).

\bibitem{Hashimoto:2009univers}
\bibinfo{author}{\bibfnamefont{K.}~\bibnamefont{Hashimoto}} \et,
\bibinfo{journal}{Phys. Rev. B} \textbf{\bibinfo{volume}{81}},
  \bibinfo{pages}{220501(R)} (\bibinfo{year}{2010}).

\bibitem{Yamashita:2009thcnP}
\bibinfo{author}{\bibfnamefont{M.}~\bibnamefont{Yamashita}} \et,
\bibinfo{journal}{Phys. Rev. B} \textbf{\bibinfo{volume}{80}},
  \bibinfo{pages}{220509(R)} (\bibinfo{year}{2009}).

\bibitem{Nakai:2010nodesP}
\bibinfo{author}{\bibfnamefont{Y.}~\bibnamefont{Nakai}} \et,
\bibinfo{journal}{Phys. Rev. B} \textbf{\bibinfo{volume}{81}},
  \bibinfo{pages}{020503(R)} (\bibinfo{year}{2010}).

\bibitem{Bendele:2010nodes11}
\bibinfo{author}{\bibfnamefont{M.}~\bibnamefont{Bendele}} \et,
\bibinfo{journal}{Phys. Rev. B} \textbf{\bibinfo{volume}{81}},
  \bibinfo{pages}{224520} (\bibinfo{year}{2010}).

\bibitem{Vekhter:1999dos}
\bibinfo{author}{\bibfnamefont{I.}~\bibnamefont{Vekhter}} \et,
\bibinfo{journal}{Phys. Rev. B} \textbf{\bibinfo{volume}{59}},
  \bibinfo{pages}{R9023} (\bibinfo{year}{1999}).

\bibitem{Vekhter:2001dTD}
\bibinfo{author}{\bibfnamefont{I.}~\bibnamefont{Vekhter}} \et,
\bibinfo{journal}{Phys. Rev. B} \textbf{\bibinfo{volume}{64}},
  \bibinfo{pages}{064513} (\bibinfo{year}{2001}).

\bibitem{Vorontsov:2006invC}
\bibinfo{author}{\bibfnamefont{A.}~\bibnamefont{Vorontsov}}
  \bibnamefont{and}
  \bibinfo{author}{\bibfnamefont{I.}~\bibnamefont{Vekhter}},
\bibinfo{journal}{Phys. Rev. Lett.} \textbf{\bibinfo{volume}{96}},
  \bibinfo{pages}{237001} (\bibinfo{year}{2006}).

\bibitem[{\citenamefont{Vorontsov and
  Vekhter}(2007{\natexlab{a}})}]{Vorontsov:2007prb1}
\bibinfo{author}{\bibfnamefont{A.}~\bibnamefont{Vorontsov}}
  \bibnamefont{and}
  \bibinfo{author}{\bibfnamefont{I.}~\bibnamefont{Vekhter}},
\bibinfo{journal}{Phys. Rev. B} \textbf{\bibinfo{volume}{75}},
  \bibinfo{pages}{224501} (\bibinfo{year}{2007}{\natexlab{a}}).

\bibitem[{\citenamefont{Vorontsov and
  Vekhter}(2007{\natexlab{b}})}]{Vorontsov:2007prb2}
\bibinfo{author}{\bibfnamefont{A.}~\bibnamefont{Vorontsov}}
  \bibnamefont{and}
  \bibinfo{author}{\bibfnamefont{I.}~\bibnamefont{Vekhter}},
\bibinfo{journal}{Phys. Rev. B} \textbf{\bibinfo{volume}{75}},
  \bibinfo{pages}{224502} (\bibinfo{year}{2007}{\natexlab{b}}).

\bibitem{Matsuda:2006review}
\bibinfo{author}{\bibfnamefont{Y.}~\bibnamefont{Matsuda}} \et,
\bibinfo{journal}{J. Phys: Cond. Mat.}
  \textbf{\bibinfo{volume}{18}}, \bibinfo{pages}{R705} (\bibinfo{year}{2006}).

\bibitem{TSakakibara:2007}
\bibinfo{author}{\bibfnamefont{T.}~\bibnamefont{Sakakibara}} \et,
\bibinfo{journal}{J. Phys. Soc. Jpn}
  \textbf{\bibinfo{volume}{76}}, \bibinfo{pages}{051004}
  (\bibinfo{year}{2007}).

\bibitem{An:2010invers}
\bibinfo{author}{\bibfnamefont{K.}~\bibnamefont{An}} \et,
\bibinfo{journal}{Phys. Rev. Lett.} \textbf{\bibinfo{volume}{104}},
  \bibinfo{pages}{037002} (\bibinfo{year}{2010}).

\bibitem{Graser:2008doppler}
\bibinfo{author}{\bibfnamefont{S.}~\bibnamefont{Graser}} \et,
\bibinfo{journal}{Phys. Rev. B}
  \textbf{\bibinfo{volume}{77}}, \bibinfo{pages}{180514(R)}
  (\bibinfo{year}{2008}).

\bibitem{Zeng:2010anisotr11}
\bibinfo{author}{\bibfnamefont{B.}~\bibnamefont{Zeng}} \et,
arXiv:1004.2236.

\bibitem{footnote}{We do not consider here possible gap modulation along the $c$-axis, see
S.~Graser et~al., Phys. Rev. B {\bf 81}, 214503 (2010) and 
J. Reid et~al., Phys. Rev. B {\bf 82}, 064501 (2010).}

\bibitem{Mishra:2009thcn}
\bibinfo{author}{\bibfnamefont{V.}~\bibnamefont{Mishra}} \et,
\bibinfo{journal}{Phys. Rev. B} \textbf{\bibinfo{volume}{80}},
  \bibinfo{pages}{224525} (\bibinfo{year}{2009}{\natexlab{a}}).

\bibitem{AHoughton:1998}
\bibinfo{author}{\bibfnamefont{A.}~\bibnamefont{Houghton}}
  \bibnamefont{and}
  \bibinfo{author}{\bibfnamefont{I.}~\bibnamefont{Vekhter}},
\bibinfo{journal}{Phys. Rev. B} \textbf{\bibinfo{volume}{57}},
  \bibinfo{pages}{10831} (\bibinfo{year}{1998}).

\bibitem{HKusunose:2004}
\bibinfo{author}{\bibfnamefont{H.}~\bibnamefont{Kusunose}},
\bibinfo{journal}{Phys. Rev. B} \textbf{\bibinfo{volume}{70}},
  \bibinfo{pages}{054509} (\bibinfo{year}{2004}).

\bibitem{Hiragi:2010para}
\bibinfo{author}{\bibfnamefont{M.}~\bibnamefont{Hiragi}} \et,
\bibinfo{journal}{J. Phys. Soc. Jpn.} \textbf{\bibinfo{volume}{79}},
  \bibinfo{pages}{094709} (\bibinfo{year}{2010}).

\bibitem{Vekhter:1999bpt}
\bibinfo{author}{\bibfnamefont{I.}~\bibnamefont{Vekhter}}
  \bibnamefont{and}
  \bibinfo{author}{\bibfnamefont{A.}~\bibnamefont{Houghton}},
\bibinfo{journal}{Phys. Rev. Lett.} \textbf{\bibinfo{volume}{83}},
  \bibinfo{pages}{4626} (\bibinfo{year}{1999}).

\bibitem{Aux} 
See supplemental EPAPS Document No. []. 

\bibitem{Boyd:2009doppler}
\bibinfo{author}{\bibfnamefont{G.~R.} \bibnamefont{Boyd}} \et,
\bibinfo{journal}{Phys. Rev. B} \textbf{\bibinfo{volume}{79}},
  \bibinfo{pages}{064525} (\bibinfo{year}{2009}).

\bibitem{Mishra:2009lift}
\bibinfo{author}{\bibfnamefont{V.}~\bibnamefont{Mishra}} \et,
\bibinfo{journal}{Phys. Rev. B} \textbf{\bibinfo{volume}{79}}, \bibinfo{pages}{094512}
  (\bibinfo{year}{2009}{\natexlab{b}}).

\bibitem{KMaki:1967}
\bibinfo{author}{\bibfnamefont{K.}~\bibnamefont{Maki}},
\bibinfo{journal}{Phys.  Rev.} \textbf{\bibinfo{volume}{158}}, \bibinfo{pages}{397}
  (\bibinfo{year}{1967}).

\bibitem{ChubukovEremin}
\bibinfo{author}{\bibfnamefont{A.V.}~\bibnamefont{Chubukov} and 
\bibfnamefont{I.}~\bibnamefont{Eremin}},
\bibinfo{journal}{Phys. Rev. B} \textbf{\bibinfo{volume}{82}}, \bibinfo{pages}{060504(R)}
  (\bibinfo{year}{2010}).

\end{thebibliography}

\begin{thebibliography}{14} \expandafter\ifx\csname natexlab\endcsname\relax\def\natexlab#1{#1}\fi
\expandafter\ifx\csname bibnamefont\endcsname\relax
  \def\bibnamefont#1{#1}\fi
\expandafter\ifx\csname bibfnamefont\endcsname\relax
  \def\bibfnamefont#1{#1}\fi
\expandafter\ifx\csname citenamefont\endcsname\relax
  \def\citenamefont#1{#1}\fi
\expandafter\ifx\csname url\endcsname\relax
  \def\url#1{\texttt{#1}}\fi
\expandafter\ifx\csname urlprefix\endcsname\relax\def\urlprefix{URL }\fi \providecommand{\bibinfo}[2]{#2}
\providecommand{\eprint}[2][]{\url{#2}}

\bibitem[{\citenamefont{Zeng et~al.}(2010)\citenamefont{Zeng, Luo, Xiang, Yang,
  Shan, Ren, Mazin, Dai, and Wen}}]{Zeng:2010anisotr11add}
\bibinfo{author}{\bibfnamefont{B.}~\bibnamefont{Zeng}},
  \bibinfo{author}{\bibfnamefont{H.~Q.} \bibnamefont{Luo}},
  \bibinfo{author}{\bibfnamefont{T.}~\bibnamefont{Xiang}},
  \bibinfo{author}{\bibfnamefont{H.}~\bibnamefont{Yang}},
  \bibinfo{author}{\bibfnamefont{L.}~\bibnamefont{Shan}},
  \bibinfo{author}{\bibfnamefont{C.}~\bibnamefont{Ren}},
  \bibinfo{author}{\bibfnamefont{I.~I.} \bibnamefont{Mazin}},
  \bibinfo{author}{\bibfnamefont{P.~C.} \bibnamefont{Dai}}, \bibnamefont{and}
  \bibinfo{author}{\bibfnamefont{H.~H.} \bibnamefont{Wen}},
  \bibinfo{journal}{arXiv} \textbf{\bibinfo{pages}{1004.2236v1}}
  (\bibinfo{year}{2010}).

\bibitem[{\citenamefont{{Borisenko} et~al.}(2010)\citenamefont{{Borisenko},
  {Zabolotnyy}, {Evtushinsky}, {Kim}, {Morozov}, {Yaresko}, {Kordyuk}, {Behr},
  {Vasiliev}, {Follath} et~al.}}]{SBorisenko:2010}
\bibinfo{author}{\bibfnamefont{S.~V.} \bibnamefont{{Borisenko}}},
  \bibinfo{author}{\bibfnamefont{V.~B.} \bibnamefont{{Zabolotnyy}}},
  \bibinfo{author}{\bibfnamefont{D.~V.} \bibnamefont{{Evtushinsky}}},
  \bibinfo{author}{\bibfnamefont{T.~K.} \bibnamefont{{Kim}}},
  \bibinfo{author}{\bibfnamefont{I.~V.} \bibnamefont{{Morozov}}},
  \bibinfo{author}{\bibfnamefont{A.~N.} \bibnamefont{{Yaresko}}},
  \bibinfo{author}{\bibfnamefont{A.~A.} \bibnamefont{{Kordyuk}}},
  \bibinfo{author}{\bibfnamefont{G.}~\bibnamefont{{Behr}}},
  \bibinfo{author}{\bibfnamefont{A.}~\bibnamefont{{Vasiliev}}},
  \bibinfo{author}{\bibfnamefont{R.}~\bibnamefont{{Follath}}},
  \bibnamefont{et~al.}, \bibinfo{journal}{ArXiv e-prints}
  (\bibinfo{year}{2010}), \eprint{1001.1147}.

\bibitem[{\citenamefont{Vorontsov and Vekhter}(2007)}]{Vorontsov:2007prb1add}
    \bibinfo{author}{\bibfnamefont{A.}~\bibnamefont{Vorontsov}} \bibnamefont{and}
  \bibinfo{author}{\bibfnamefont{I.}~\bibnamefont{Vekhter}},
  \bibinfo{journal}{Phys. Rev. B} \textbf{\bibinfo{volume}{75}},
  \bibinfo{pages}{224501} (\bibinfo{year}{2007}).

\bibitem[{\citenamefont{Graser et~al.}(2010)\citenamefont{Graser, Kemper,
  Maier, Cheng, Hirschfeld, and Scalapino}}]{SGraser:2010}
\bibinfo{author}{\bibfnamefont{S.}~\bibnamefont{Graser}},
  \bibinfo{author}{\bibfnamefont{A.~F.} \bibnamefont{Kemper}},
  \bibinfo{author}{\bibfnamefont{T.~A.} \bibnamefont{Maier}},
  \bibinfo{author}{\bibfnamefont{H.-P.} \bibnamefont{Cheng}},
  \bibinfo{author}{\bibfnamefont{P.~J.} \bibnamefont{Hirschfeld}},
  \bibnamefont{and} \bibinfo{author}{\bibfnamefont{D.~J.}
  \bibnamefont{Scalapino}}, \bibinfo{journal}{Phys. Rev. B}
  \textbf{\bibinfo{volume}{81}}, \bibinfo{pages}{214503}
  (\bibinfo{year}{2010}).

\bibitem[{\citenamefont{{Reid} et~al.}(2010)\citenamefont{{Reid}, {Tanatar},
  {Luo}, {Shakeripour}, {Doiron-Leyraud}, {Ni}, {Bud'ko}, {Canfield},
  {Prozorov}, and {Taillefer}}}]{JReid:2010}
\bibinfo{author}{\bibfnamefont{J.}~\bibnamefont{{Reid}}},
  \bibinfo{author}{\bibfnamefont{M.~A.} \bibnamefont{{Tanatar}}},
  \bibinfo{author}{\bibfnamefont{X.~G.} \bibnamefont{{Luo}}},
  \bibinfo{author}{\bibfnamefont{H.}~\bibnamefont{{Shakeripour}}},
  \bibinfo{author}{\bibfnamefont{N.}~\bibnamefont{{Doiron-Leyraud}}},
  \bibinfo{author}{\bibfnamefont{N.}~\bibnamefont{{Ni}}},
  \bibinfo{author}{\bibfnamefont{S.~L.} \bibnamefont{{Bud'ko}}},
  \bibinfo{author}{\bibfnamefont{P.~C.} \bibnamefont{{Canfield}}},
  \bibinfo{author}{\bibfnamefont{R.}~\bibnamefont{{Prozorov}}},
  \bibnamefont{and}
  \bibinfo{author}{\bibfnamefont{L.}~\bibnamefont{{Taillefer}}},
 \bibinfo{journal}{Phys. Rev. B} \textbf{\bibinfo{volume}{82}}, \bibinfo{pages}{064501}
  (\bibinfo{year}{2010}).

\bibitem[{\citenamefont{Nakayama et~al.}(2009)\citenamefont{Nakayama, Sato,
  Richard, Xu, Sekiba, Souma, Chen, Luo, Wang, Ding
  et~al.}}]{Nakayama:2009arpesK122add}
\bibinfo{author}{\bibfnamefont{K.}~\bibnamefont{Nakayama}},
  \bibinfo{author}{\bibfnamefont{T.}~\bibnamefont{Sato}},
  \bibinfo{author}{\bibfnamefont{P.}~\bibnamefont{Richard}},
  \bibinfo{author}{\bibfnamefont{Y.}~\bibnamefont{Xu}},
  \bibinfo{author}{\bibfnamefont{Y.}~\bibnamefont{Sekiba}},
  \bibinfo{author}{\bibfnamefont{S.}~\bibnamefont{Souma}},
  \bibinfo{author}{\bibfnamefont{G.}~\bibnamefont{Chen}},
  \bibinfo{author}{\bibfnamefont{J.}~\bibnamefont{Luo}},
  \bibinfo{author}{\bibfnamefont{N.}~\bibnamefont{Wang}},
  \bibinfo{author}{\bibfnamefont{H.}~\bibnamefont{Ding}}, \bibnamefont{et~al.},
  \bibinfo{journal}{Europhys. Lett.} \textbf{\bibinfo{volume}{85}},
  \bibinfo{pages}{67002} (\bibinfo{year}{2009}).

\bibitem[{\citenamefont{Ding et~al.}(2008)\citenamefont{Ding, Richard,
  Nakayama, Sugawara, Arakane, Sekiba, Takayama, Souma, Sato, Takahashi
  et~al.}}]{Ding:2008arpesK122add}
\bibinfo{author}{\bibfnamefont{H.}~\bibnamefont{Ding}},
  \bibinfo{author}{\bibfnamefont{P.}~\bibnamefont{Richard}},
  \bibinfo{author}{\bibfnamefont{K.}~\bibnamefont{Nakayama}},
  \bibinfo{author}{\bibfnamefont{K.}~\bibnamefont{Sugawara}},
  \bibinfo{author}{\bibfnamefont{T.}~\bibnamefont{Arakane}},
  \bibinfo{author}{\bibfnamefont{Y.}~\bibnamefont{Sekiba}},
  \bibinfo{author}{\bibfnamefont{A.}~\bibnamefont{Takayama}},
  \bibinfo{author}{\bibfnamefont{S.}~\bibnamefont{Souma}},
  \bibinfo{author}{\bibfnamefont{T.}~\bibnamefont{Sato}},
  \bibinfo{author}{\bibfnamefont{T.}~\bibnamefont{Takahashi}},
  \bibnamefont{et~al.}, \bibinfo{journal}{Europhys. Lett.}
  \textbf{\bibinfo{volume}{83}}, \bibinfo{pages}{47001} (\bibinfo{year}{2008}).

\bibitem[{\citenamefont{Udagawa et~al.}(2005)\citenamefont{Udagawa, Yanase, and
  Ogata}}]{MUdagawa:2005}
\bibinfo{author}{\bibfnamefont{M.}~\bibnamefont{Udagawa}},
  \bibinfo{author}{\bibfnamefont{Y.}~\bibnamefont{Yanase}}, \bibnamefont{and}
  \bibinfo{author}{\bibfnamefont{M.}~\bibnamefont{Ogata}},
  \bibinfo{journal}{Phys. Rev. B} \textbf{\bibinfo{volume}{71}},
  \bibinfo{pages}{024511} (\bibinfo{year}{2005}).

\bibitem[{\citenamefont{Nagai et~al.}(2007)\citenamefont{Nagai, Kato, Hayashi,
  Yamauchi, and Harima}}]{YNagai:2007}
\bibinfo{author}{\bibfnamefont{Y.}~\bibnamefont{Nagai}},
  \bibinfo{author}{\bibfnamefont{Y.}~\bibnamefont{Kato}},
  \bibinfo{author}{\bibfnamefont{N.}~\bibnamefont{Hayashi}},
  \bibinfo{author}{\bibfnamefont{K.}~\bibnamefont{Yamauchi}}, \bibnamefont{and}
  \bibinfo{author}{\bibfnamefont{H.}~\bibnamefont{Harima}},
  \bibinfo{journal}{Phys. Rev. B} \textbf{\bibinfo{volume}{76}},
  \bibinfo{pages}{214514} (\bibinfo{year}{2007}).

\bibitem[{\citenamefont{Nagai et~al.}(2009)\citenamefont{Nagai, Hayashi, Kato,
  Yamauchi, and Harima}}]{YNagai:2009}
\bibinfo{author}{\bibfnamefont{Y.}~\bibnamefont{Nagai}},
  \bibinfo{author}{\bibfnamefont{N.}~\bibnamefont{Hayashi}},
  \bibinfo{author}{\bibfnamefont{Y.}~\bibnamefont{Kato}},
  \bibinfo{author}{\bibfnamefont{K.}~\bibnamefont{Yamauchi}}, \bibnamefont{and}
  \bibinfo{author}{\bibfnamefont{H.}~\bibnamefont{Harima}},
  \bibinfo{journal}{Journal of Physics: Conference Series}
  \textbf{\bibinfo{volume}{150}}, \bibinfo{pages}{052177}
  (\bibinfo{year}{2009}).

\bibitem[{\citenamefont{Vekhter and Vorontsov}(2008)}]{IVekhter:2008}
    \bibinfo{author}{\bibfnamefont{I.}~\bibnamefont{Vekhter}} \bibnamefont{and}
  \bibinfo{author}{\bibfnamefont{A.}~\bibnamefont{Vorontsov}},
  \bibinfo{journal}{Physica B: Condensed Matter}
  \textbf{\bibinfo{volume}{403}}, \bibinfo{pages}{958 } (\bibinfo{year}{2008}).

\bibitem[{\citenamefont{Shishido et~al.}(2010)\citenamefont{Shishido, Bangura,
  Coldea, Tonegawa, Hashimoto, Kasahara, Rourke, Ikeda, Terashima, Settai
  et~al.}}]{HShishido:2010}
\bibinfo{author}{\bibfnamefont{H.}~\bibnamefont{Shishido}},
  \bibinfo{author}{\bibfnamefont{A.~F.} \bibnamefont{Bangura}},
  \bibinfo{author}{\bibfnamefont{A.~I.} \bibnamefont{Coldea}},
  \bibinfo{author}{\bibfnamefont{S.}~\bibnamefont{Tonegawa}},
  \bibinfo{author}{\bibfnamefont{K.}~\bibnamefont{Hashimoto}},
  \bibinfo{author}{\bibfnamefont{S.}~\bibnamefont{Kasahara}},
  \bibinfo{author}{\bibfnamefont{P.~M.~C.} \bibnamefont{Rourke}},
  \bibinfo{author}{\bibfnamefont{H.}~\bibnamefont{Ikeda}},
  \bibinfo{author}{\bibfnamefont{T.}~\bibnamefont{Terashima}},
  \bibinfo{author}{\bibfnamefont{R.}~\bibnamefont{Settai}},
  \bibnamefont{et~al.}, \bibinfo{journal}{Phys. Rev. Lett.}
  \textbf{\bibinfo{volume}{104}}, \bibinfo{pages}{057008}
  (\bibinfo{year}{2010}).

\bibitem[{\citenamefont{Sato et~al.}(2009)\citenamefont{Sato, Nakayama, Sekiba,
  Richard, Xu, Souma, Takahashi, Chen, Luo, Wang et~al.}}]{Sato:2009p589}
\bibinfo{author}{\bibfnamefont{T.}~\bibnamefont{Sato}},
  \bibinfo{author}{\bibfnamefont{K.}~\bibnamefont{Nakayama}},
  \bibinfo{author}{\bibfnamefont{Y.}~\bibnamefont{Sekiba}},
  \bibinfo{author}{\bibfnamefont{P.}~\bibnamefont{Richard}},
  \bibinfo{author}{\bibfnamefont{Y.~M.} \bibnamefont{Xu}},
  \bibinfo{author}{\bibfnamefont{S.}~\bibnamefont{Souma}},
  \bibinfo{author}{\bibfnamefont{T.}~\bibnamefont{Takahashi}},
  \bibinfo{author}{\bibfnamefont{G.~F.} \bibnamefont{Chen}},
  \bibinfo{author}{\bibfnamefont{J.~L.} \bibnamefont{Luo}},
  \bibinfo{author}{\bibfnamefont{N.~L.} \bibnamefont{Wang}},
  \bibnamefont{et~al.}, \bibinfo{journal}{Phys. Rev. Lett.}
  \textbf{\bibinfo{volume}{103}}, \bibinfo{pages}{047002}
  (\bibinfo{year}{2009}).

\bibitem[{\citenamefont{Zabolotnyy et~al.}(2009)\citenamefont{Zabolotnyy,
  Inosov, Evtushinsky, Koitzsch, Kordyuk, Sun, Park, Haug, Hinkov, Boris
  et~al.}}]{Zabolotnyy:2009eorder}
\bibinfo{author}{\bibfnamefont{V.~B.} \bibnamefont{Zabolotnyy}},
  \bibinfo{author}{\bibfnamefont{D.~S.} \bibnamefont{Inosov}},
  \bibinfo{author}{\bibfnamefont{D.~V.} \bibnamefont{Evtushinsky}},
  \bibinfo{author}{\bibfnamefont{A.}~\bibnamefont{Koitzsch}},
  \bibinfo{author}{\bibfnamefont{A.~A.} \bibnamefont{Kordyuk}},
  \bibinfo{author}{\bibfnamefont{G.~L.} \bibnamefont{Sun}},
  \bibinfo{author}{\bibfnamefont{J.~T.} \bibnamefont{Park}},
  \bibinfo{author}{\bibfnamefont{D.}~\bibnamefont{Haug}},
  \bibinfo{author}{\bibfnamefont{V.}~\bibnamefont{Hinkov}},
  \bibinfo{author}{\bibfnamefont{A.~V.} \bibnamefont{Boris}},
  \bibnamefont{et~al.}, \bibinfo{journal}{Nature}
  \textbf{\bibinfo{volume}{457}}, \bibinfo{pages}{569} (\bibinfo{year}{2009}).

\end{thebibliography}


\end{document}